\documentclass[aps,prl,twocolumn, titlepage,showpacs]{revtex4}

\usepackage{graphicx}

\usepackage{dcolumn}

\usepackage{bm}

\begin{document}

\title{Viscoelasticity of 2D liquids quantified in a dusty plasma experiment}

\author{Yan Feng}
\email{yan-feng@uiowa.edu}
\author{J. Goree}
\author{Bin Liu}
\affiliation{Department of Physics and Astronomy, The University
of Iowa, Iowa City, Iowa 52242}

\date{\today}

\begin{abstract}

The viscoelasticity of two-dimensional liquids is quantified in an
experiment using a dusty plasma. An experimental method is
demonstrated for measuring the wavenumber-dependent viscosity,
$\eta(k)$, which is a quantitative indicator of viscoelasticity.
Using an expression generalized here to include friction,
$\eta(k)$ is computed from the transverse current autocorrelation
function (TCAF), which is found by tracking random particle
motion. The TCAF exhibits an oscillation that is a signature of
elastic contributions to viscoelasticity. Simulations of a Yukawa
liquid are consistent with the experiment.

\end{abstract}

\narrowtext

\maketitle

Two-dimensional (2D) physical systems include electrons on a
liquid helium surface~\cite{Grimes:79}, colloids~\cite{Murray:90},
granular fluids~\cite{Res:06}, and dusty
plasmas~\cite{Morfill:09}. In experiments and simulations, elastic
properties, such as transverse waves~\cite{Nunomura:00, Donko:04},
and transport properties, such as viscosity
$\eta$~\cite{Nosenko:04, Liu:05, Donko:06}, have been studied.

Viscoelasticity is a property of materials that exhibit both
viscous and elastic characteristics~\cite{Lakes:09}. One usually
thinks of viscous properties for liquids and elastic properties
for solids, but most materials are viscoelastic and exhibit both.
These include, for example polymers, human tissue, and hot
metal~\cite{Lakes:09}. In general, liquids exhibit elastic effects
especially at short length or time scales~\cite{Nosenko:06_2}, but
viscous effects at long length or time scales.

To quantify viscoelasticity, one often uses the
frequency-dependent viscosity $\eta(\omega)$~\cite{Mason:95},
which tends toward the static viscosity, $\eta$, as $\omega
\rightarrow 0$. The $\eta(\omega)$ is easily measured in
three-dimensional (3D) liquids using rheometers and
viscometers~\cite{Mason:95}, but not in most 2D liquids.

Besides $\eta(\omega)$, the wavenumber-dependent viscosity,
$\eta(k)$, has been used by theorists to quantify the viscoelastic
character~\cite{Alley:85, Balucani:87, Palmer:94, Balucani:00,
Hu:07}. They have recently developed ways of computing $\eta(k)$
from the trajectories of random motion of
molecules~\cite{Balucani:00, Hu:07}. However, until now, there
have been no experimental measurements in any physical systems
known to us of $\eta(k)$ that exploit this new analysis method.
One difficulty in using this method in an experiment is that it
requires, as its inputs, the positions $x_i$ and velocities $v_i$
of $N$ individual molecules or particles as they move about
randomly. In this Letter, we will use an experimental system,
dusty plasma, that allows observing these inputs directly.

Here we further develop a method for computing $\eta(k)$,
generalizing it for multiphase systems like dusty plasma. As was
originally developed for 3D molecular dynamics (MD) simulations,
the method begins with computing the normalized transverse current
autocorrelation function~\cite{Balucani:00, Hu:07} (TCAF), which
is defined as
\begin{equation}\label{TCAF}
{C_T(k,t) = \langle j^*_y(k,0)\,j_y(k,t)\rangle / \langle
j^*_y(k,0)\,j_y(k,0)\rangle ,}
\end{equation}
where $j_y(k,t)=\sum\nolimits^N_{i=1} v^y_i(t)\,{\rm
exp}[ikx_i(t)]$ is the transverse current, with the vector $k$
parallel to the $x$ axis. Then, $\eta(k)$ can be
calculated~\cite{Balucani:00, Hu:07} through
$\eta(k)/\rho=1/(k^2\,\Phi)$, where $\Phi$ is the area under the
normalized TCAF. This equation can be derived, assuming that the
viscosity is a valid transport coefficient, either from the
hydrodynamic Navier-Stokes equation or from linear response
theory~\cite{Hu:07}. Here, we generalize this equation using
${\partial \mathbf{j}(\mathbf{r},t)}/{\partial t}-(\eta
/\rho)\mathbf{\nabla}^2 \mathbf{j}(\mathbf{r},t)+\nu_f
\mathbf{j}(\mathbf{r},t)=0$, a Navier-Stokes equation that
includes an additional frictional drag force $\nu_f
\mathbf{j}(\mathbf{r},t)$ due to a second phase \cite{Nosenko:04}.
This equation is valid in both 2D and 3D systems. Following the
method of \cite{Hu:07}, we find~\cite{derive}
\begin{equation}\label{eta}
{\eta(k)/\rho=((1/\Phi)-\nu_f)/k^2.}
\end{equation}

Here we will characterize viscoelasticity in an experiment two
ways. First, as a signature of elastic effects, we will detect
oscillations in the TCAF~\cite{Balucani:87, Balucani:00} for large
$k$. Second, we will measure the diminishment of $\eta(k)$ as $k$
increases. This diminishment occurs along with a relative increase
of elastic contributions to viscoelasticity, for large $k$.

Dusty (complex) plasma, is partially ionized gas containing
micron-size particles of solid matter~\cite{Morfill:09, Melzer:96,
Feng:08}. Particles have a charge $Q$ and can be electrically
confined in a single horizontal layer where they self-organize
with a structure like a crystalline solid~\cite{Feng:08}. Coulomb
repulsion is shielded with a screening length
$\lambda_D$~\cite{Konopka:00, Vaulina:08}. The elastic properties
of the crystalline solid arise from interparticle repulsion and
can be characterized by the phonon spectrum for longitudinal and
transverse waves~\cite{Nunomura:05}, which have a frequency close
to the nominal 2D dust plasma frequency
$\omega_{pd}$~\cite{Kalman:04}. The solid can be melted, to form a
liquid, by laser manipulation~\cite{Nosenko:06, Wolter:05}.

Dusty plasmas are attractive for experimental quantification of
viscoelastic effects at a microscopic scale. As in
colloids~\cite{Murray:90} and granular fluids~\cite{Res:06}, they
allow video microscopy to track the $x_i$ and $v_i$ of individual
particles. They also provide both elastic and viscous effects. The
particles are immersed in a medium that is a rarefied gas that
does not overdamp particle motion, unlike
colloids~\cite{Murray:90} with their solvents.

Dusty plasma experiments, until now, have yielded descriptive
presentations of viscoelasticity~\cite{Ratynskaia:06} and
demonstrations of the microscopic motion of particles associated
with viscoelastic response~\cite{Chan:07}. In experiments, the
static viscosity has been measured~\cite{Nosenko:04} and estimated
from diffusion observations~\cite{Konopka:00_2}. However, a
quantitative characterization of viscoelasticity, using
$\eta(\omega)$ or $\eta(k)$, is lacking from the literature.

A challenge in dusty plasma experiments is that they do not allow
direct contact of the suspension with a container. Thus, the
viscoelastic response cannot be measured with a rheometer. We
overcome this challenge by observing the random particle motion
and using Eq.~(\ref{eta}) to compute $\eta(k)$. We will do this
with experimental data, and confirm our interpretation using a
simulation.

Using the apparatus of \cite{Liu:08}, a plasma was powered by
$13.56~\rm{MHz}$, $170~\rm{V}$ peak to peak voltages. After the
$8.1~\rm{\mu m}$ diameter microspheres were introduced into the
plasma (which had an Argon pressure of $14~\rm{mTorr}$), they
experienced a damping rate of
$\nu_f=2.4~\rm{s^{-1}}$~\cite{Liu:03}.

The particles were suspended in a single layer. They
self-organized in a triangular lattice~\cite{Feng:08}. Particle
motion was essentially 2D, with negligible out-of-plane
displacements. The suspension had a diameter $\approx 52~\rm{mm}$
and contained $> 5400$ particles. The lattice constant
$b=0.67~\rm{mm}$ corresponds to a Wigner-Seitz
radius~\cite{Kalman:04} $a = 0.35~\rm{mm}$.

Particle tracking was done by imaging from the top. For each of
four runs, $20~\rm{s}$ videos were recorded at
$250~{\rm{frames/s}}$, providing adequate time resolution for the
TCAF. The $(36.2 \times 22.6)~\rm{mm^2}$ field of view (FOV)
included $\approx 2100$ particles. We recorded the maximum 5061
frames per run allowed by the 12-bit Phantom v5.2 camera, with a
lens that provided a resolution of 0.03~mm/pixel. For each video
frame $j$, we computed~\cite{Feng:07} the position of the
{\it{i}}th particle, $\tilde{x}_{i,j}$. To compute $j_y(k,t)$, we
used $x_{i,j}=(\tilde{x}_{i,
j-1}+\tilde{x}_{i,j}+\tilde{x}_{i,j+1})/3$ and
$v^y_{i,j}=(\tilde{y}_{i,j+1}-\tilde{y}_{i,j-1})/2 \delta t$. This
finite-difference method reduced errors arising from the high
frame rate. Examples of particle trajectories from the experiment
are shown in Fig.~1(a). Next, we computed $j_y(k,t)$ and smoothed
its time series over five frames before calculating the TCAF,
Eq.~(\ref{TCAF}), and finally $\eta(k)$, Eq.~(\ref{eta}).

Before melting the suspension, we used the phonon-spectrum method
for a lattice~\cite{Nunomura:05} to measure $Q/e=-6000$,
$\kappa_0=a/\lambda_D=0.5$, and $\omega_{pd}=30~\rm{s^{-1}}$.
After melting, we determined $T$ from the mean-square velocity
fluctuation~\cite{Feng:08} yielding
$\Gamma=(Q^2/4\pi\epsilon_0a)/(k_BT)=68$.

We melted the lattice and maintained a steady kinetic temperature
$T$ using laser manipulation~\cite{Nosenko:06, Liu:08}. Random
kicks were applied by radiation pressure from a pair of 532-nm
laser beams that were rastered across the suspension in a
Lissajous pattern with frequencies $f_x=48.541$~Hz and
$f_y=30$~Hz. This pattern filled a rectangle larger than the
camera's FOV. Along with the desired random motion, the Lissajous
heating method also produces coherent modes~\cite{Nosenko:06},
which had about 8\% of our total kinetic energy for motion in the
$y$ direction, similar to~\cite{Liu:08}. We analyzed half of the
FOV, where the temperature was uniform within extremes of $\pm
20\%$.

For comparison to the experiment, we also performed a Langevin MD
simulation~\cite{Feng:08_2, Hou:09, Vaulina:09, Ott:09} of a 2D
Yukawa liquid to mimic our experiment. Using periodic boundary
conditions and 4096 particles, the equation of motion Eq.~(3)
of~\cite{Feng:08_2} was integrated, yielding particle
trajectories, Fig.~1(b). The simulation parameters $\Gamma = 68$,
$\kappa_0 = 0.5$, and $\nu_f/\omega_{pd} = 0.08$ match the
experimental values. To improve statistics, the simulation was run
much longer, $\omega_{pd}t=22~300$, than the experiment
$\omega_{pd}t=607$. To validate our Langevin MD simulation, we
also performed a frictionless MD simulation~\cite{Liu:05} and
calculated $\eta(k)$ as in Eq.(\ref{eta}) but with $\nu_f = 0$; we
found that the results for $\eta(k)$ for the two types of
simulations agree. In addition to computing $\eta(k)$, we also
computed the static viscosity $\eta$ using the Green-Kubo
relation, Eq.~(3) of~\cite{Liu:05}. The latter assumes that the
shear-stress autocorrelation function decays significantly faster
than $1/t$, which we verified.

Experimental results for the TCAF, Fig.~2(a), reveal elastic
properties in the viscoelastic regime for this liquid. The TCAF
computed from Eq.~(\ref{TCAF}) exhibits an initial decay followed
by oscillations around zero~\cite{Balucani:00, Balucani:87}, for
$kb=3.26$ in Fig.~2(a). Such oscillations typically indicate that
the selected wavenumber corresponds to the viscoelastic regime.
The TCAF is a time series; we also calculate its frequency
spectrum, shown in the inset of Fig.~2(a). (This frequency
spectrum can also be used in generating a phonon
spectrum~\cite{Nosenko:06_2}). The spectrum features a prominent
peak at non-zero frequency. This peak is a signature of shear
elasticity; it would be absent in a viscous regime. To our
knowledge, the TCAF time series has not previously been reported
for dusty plasma experiments as an indicator of viscoelasticity.

Simulation results, Fig.~2(b), exhibit features in the TCAF and
its spectrum~\cite{Balucani:87} similar to those in the
experiment. This agreement between experiment and simulation lends
confidence to our use of the TCAF as a quantitative indicator of
viscoelasticity in an experimental system.

For wavenumbers much smaller than those shown in Fig.~2, i.e., for
very long wavelengths, we would expect viscous behavior
characterized by a simple decay of the TCAF with no oscillations.
This hydrodynamic regime has been well studied in simulations and
theory~\cite{Hansen:86}. Observing it requires a sufficiently
large system. One of the attractions of our physical system is
that it allows direct observation of motion at an atomistic scale.
Thus, we use it here to observe the viscoelastic regime (at small
wavelengths), not the purely viscous hydrodynamic regime.

As our chief result, our experimentally measured
wavenumber-dependent viscosity, $\eta(k)$, is presented
quantitatively in Fig.~3(a). We observe that $\eta(k)$ diminishes
as $k$ increases. Physically, this trend indicates that
dissipative or viscous effects diminish at shorter length scales.
At these shorter length scales, elasticity has a greater effect.

Since previous experiments are not available for quantitative
comparison, we compare our experimental results to the Langevin
simulation, Fig.~3(b). We note that $\eta(k)$ exhibits the same
downward trend and similar quantitative values in the experiment
and the simulations. For both the experiment and simulation, we
present results for $\eta(k)$, computed using Eq.~(\ref{eta}), for
the viscoelastic regime, i.e., $k> 1/b$. For each $k$, the
infinite time limit for the integration of $\Phi$ was replaced
with $t_I$, the time of the first upward zero-crossing of TCAF
time series (Fig.~2). This integration limit retains both the
viscous effects at short time and the elastic effects within the
first negative peak.

Noise in the experimental results arose from the finite amount of
current data used to compute the TCAF. To verify that this
accounts for the scatter in the experimental $\eta(k)$ in
Fig.~3(a), we repeated the simulation with a shorter time,
matching the experiment not only in duration but also in particle
number. This test shows, in Fig.~3(b), that scatter arises from
the finiteness of the $j_y(k,t)$ data to the same extent as in the
experiment. In both the experiment and in the shorter simulation,
a few TCAF curves were too noisy to analyze, with a lack of a
well-defined upward zero-crossing; the corresponding few data
points are omitted from Fig.~3.

We fit $\eta(k)$ in Fig.~3 to the same empirical Pad\'e
approximant used originally for MD simulations of 3D liquids of
hard spheres~\cite{Alley:85} and water~\cite{Balucani:00}. This
approximant, $\eta(k) \propto (1+\alpha k^2)^{-1}$, apparently has
never been applied for 2D liquids. We found that this form fits
both our experimental and simulation data in Fig.~3 as well as the
scatter allows. However, a simple power law does not fit the
$\eta(k)$ data as well.

In addition to finding that our $\eta(k)$ fits the Pad\'e
approximant, we also find in Fig.~3(b), that it extrapolates as $k
\rightarrow 0$ to the static viscosity $\eta$~\cite{Balucani:00}.
In this test, we found $\eta$ using the Green-Kubo
relation~\cite{Liu:05} with our Langevin simulation; and this
result, shown as a star in Fig.~3(b), agrees with previous
simulations that used different methods~\cite{Liu:05, Donko:06}.

In conclusion, we performed an experiment to quantify
viscoelasticity of 2D liquids using the TCAF and $\eta(k)$. We did
this using measurements of random particle motion in a dusty
plasma, which is a frictional system. We generalized a method of
calculating $\eta(k)$ by including the friction in the
Navier-Stokes equation; and we presented an experimental
demonstration of this method. Our experimental results for
$\eta(k)$ show that it diminishes with increasing $k$ that can be
modeled as $\propto (1+\alpha k^2)^{-1}$, which compares well with
simulation results.

This work was supported by NSF and NASA. We thank Zhonghan Hu for
helpful discussions.

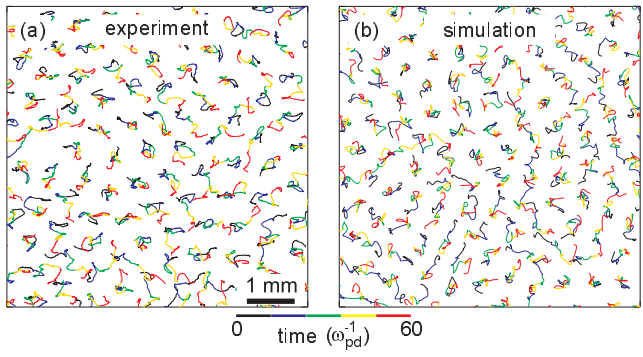
\begin{figure}[p]
\caption{\label{random motion} (Color online) Particle
trajectories in a 2D liquid, with color representing time. To
illustrate the random particle motion, (a) shows $\approx 10\%$ of
the spatial region we analyzed, for a duration
$60~\omega_{pd}^{-1}$ which is about $\approx 10\%$ of one movie,
i.e., one run in the experiment, while (b) is a part of a Langevin
MD simulation, shown over the same time interval.}
\end{figure}

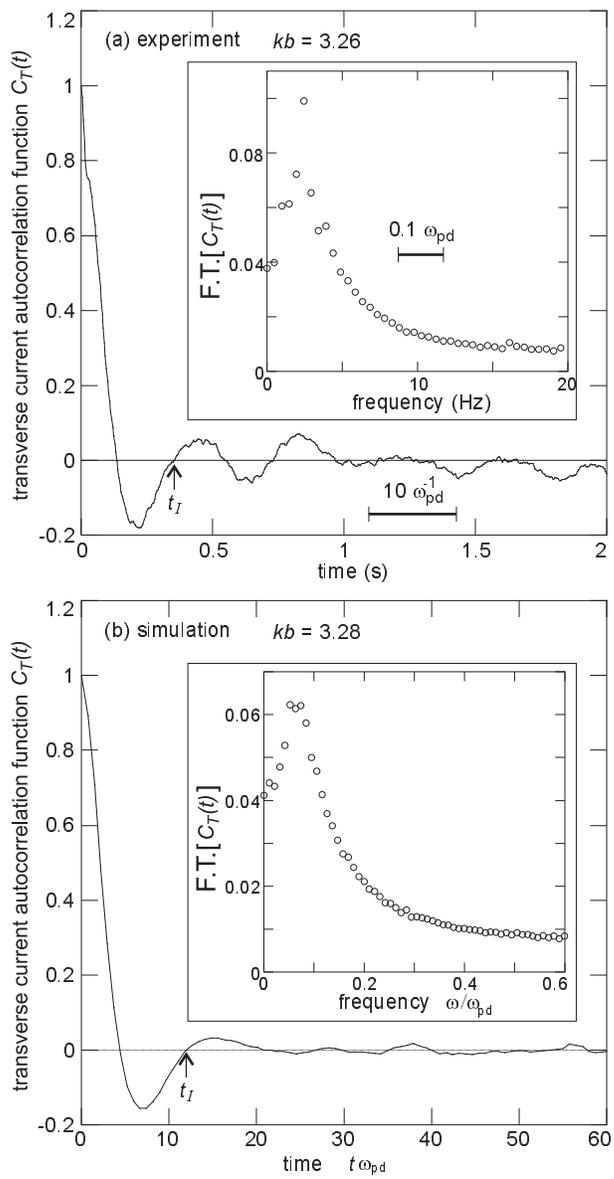
\begin{figure}[p]
\caption{\label{TCAFcurve} Transverse current autocorrelation
function (TCAF) in the 2D liquid computed using Eq.~(\ref{TCAF})
for (a) the experiment at $kb=3.26$, and (b) the Langevin MD
simulation at $kb=3.28$. At short times, the TCAF decays due to
viscous effects, while at longer times (after its first positive
zero crossing, $t_I$) it oscillates due to elastic effects. The
frequency spectrum for each TCAF, shown in the insets, reveals a
peak that is a signature of the elastic contribution to
viscoelasticity. These results are different from the pure
monotonic decay of TCAF and its spectrum that would be observed in
a purely viscous regime. (Here, $b$ is the lattice constant
measured before melting.)}
\end{figure}

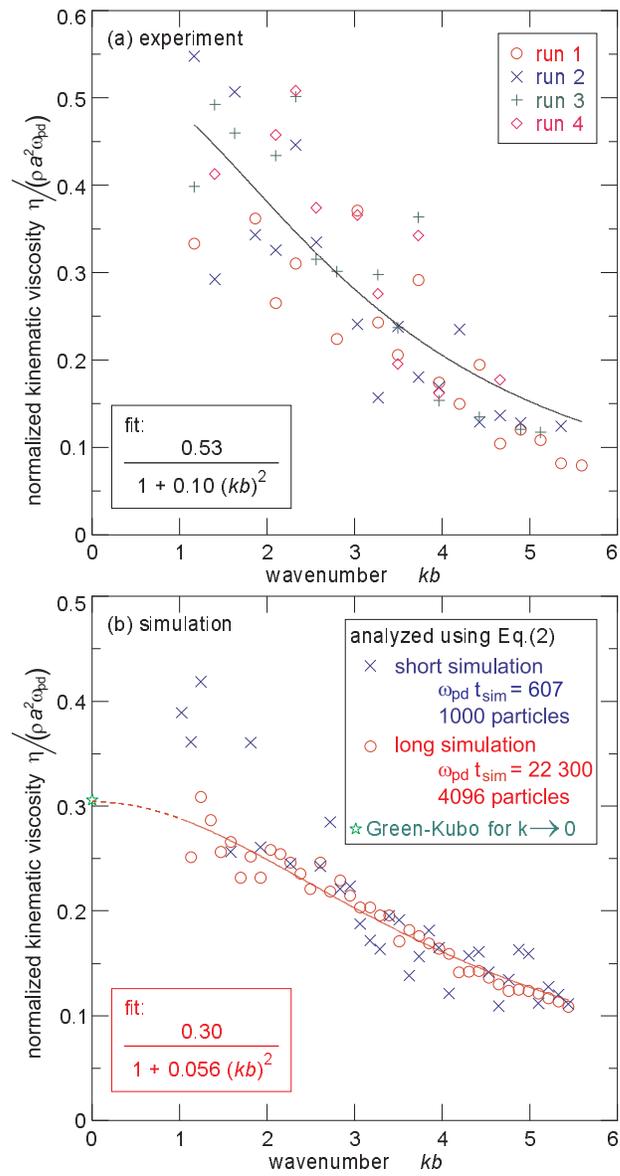
\begin{figure}[p]
\caption{\label{viscosity} (Color online) The wavenumber-dependent
viscosity $\eta(k)$ of the 2D liquid, computed using
Eq.~(\ref{eta}) for (a) the experiment and (b) simulations of two
sizes. We find that $\eta(k)$ diminishes with $k$, which is a
signature of viscoelastic effects. The size of the smaller
simulation mimics the size of the experiment; comparing them
reveals that the scatter of the experimental data (a) arises from
the data size. In (b), the Green-Kubo (static) viscosity $\eta$ is
indicated by a star symbol. Here, the kinematic viscosity
$\eta(k)/\rho$ and wavenumber $k$ are normalized to be
dimensionless.}
\end{figure}

\end{document}